# An Efficient 3D Indoor Positioning System Based on Visible Light Communication

Vailet Hikmat Faraj Al Khattat
Wireless and Photonics Networks Research Center (WiPNET), Department of Computer and Communication Systems Engineering, Faculty of Engineering
Universiti Putra Malaysia, 43400 UPM Serdang, Selangor, Malaysia
eng.vailet@gmail.com

Siti Barirah Ahmad Anas
Wireless and Photonics Networks Research Center (WiPNET), Department of Computer and Communication Systems Engineering, Faculty of Engineering
Universiti Putra Malaysia, 43400 UPM Serdang, Selangor, Malaysia
barirah@upm.edu.my

Abdu Saif
Department of Electrical Engineering, Faculty of Engineering, University of Malaya, 50603, Kuala Lumpur, Malaysia
saif.abduh2016@gmail.com

*Abstract*— In this communication revolution era, visible light communication (VLC) is the optimum efficacious answer to the increased request for high-speed data transmission with reduced cost, besides the illumination. This technology is considered the best promising candidate for indoor positioning due to its ability to attain higher positioning accuracy than other technologies, alongside its distinct advantages. However, the existing techniques that had been applied in an indoor positioning system (IPS) based on VLC focused on the aspect of accuracy but neglected the aspects of cost and complexity. This paper proposes a new technique for positioning, namely complementary and supplementary angles based on received signal strength (CSA-RSS). The positioning error performance of the system is examined based on the single light-emitting diode (LED) and different positions of the photodiode (PD). The results demonstrate the improved accuracy and prove the effectiveness of the newly proposed technique where the average error is 3.2 cm in 80% of the examined positions and the maximum average positioning error is 4.2 cm. Moreover, the proposed approach has been proven to increase system efficiency which acts as a suitable VLC design for indoor positioning performance, avoiding weak signals and poor accuracy.

*Keywords—Indoor positioning, VLC, CSA-RSS, LED, PD.*

## I. Introduction

With the rapid development of new technologies in wireless sensor networks and mobile services, visible light communication (VLC) has significantly expanded in recent years [1]. Future indoor positioning (IP) applications show significant promise with the VLC system, especially when aiming for high positioning accuracy. Using a global positioning system in indoor environments is limited by weak signals and poor accuracy in huge complexes such as public institutions, museums, underground parking garages, and other indoor situations [2]. A significant interest emerged to obtain a high-accuracy indoor positioning system (IPS) where it is fundamental to the real-time tracking of devices, people, and tools [3]. Much research had been done in the IP domain due to its benefits in many scenarios, and applications, like industry fields, health sectors, navigation, and indoor general locations. Many technologies based on IP have been developed over the last decades such as wireless fidelity (Wi-Fi), Bluetooth, ultrasound, infrared, radio frequency identification (RFID), and radio frequency (RF)_based techniques [4]. These technologies can attain at the centimeter-to-meter level in indoor positioning. However, they are constrained in terms of electromagnetic interference, complexity, low security, and expensive deployment because of needed costs and additional infrastructure [5]. The positioning precision and cost represent a tolerance issue in wireless positioning technologies [6]. Besides that, these RF-based systems may not be appropriate in places where RF is restricted like hospitals, laboratories, and airplanes because of the interference of the RF [7]. In order to overcome these restrictions, VLC technology appeared as a promising green solution providing high-speed communication applications besides meeting the needs of daily lighting [8]. Moreover, VLC became an interesting technology as the interference-free alternative to existing radio frequency communication technologies due to its significant features such as free licensing, and cost-efficiency. High precision and security can be achieved where VLC cannot penetrate the walls [9].

Artificial lighting has become widespread everywhere, from usual light bulbs on ceilings to headlights of a car. It is predicted that by 2025 light-emitting diodes (LEDs) will represent about 98% of the lighting. This attracts more interest in VLC technology in which LEDs are utilized for illumination and data transmission [10].

In this context, several techniques have been investigated and applied in VLC positioning systems. The techniques that are based on trilateration and triangulation are paid the most attention due to good accuracy if compared to the others. These techniques are the angle of arrival (AOA), time of arrival (TOA), time difference of arrival (TDOA), and received signal strength (RSS). The AOA technique can be applied to find the target by finding the intersection of the angle direction lines of the signal that is measured from the reference signals of transmitters (TXs) and the receiver (RX) [11]. The TOA technique computes the required arrival time of the signal from several TXs to the RX; the estimation is calculated based on the intersections of TXs circles. The TDOA technique calculates the difference between the arrival time between signals from at least three TXs to estimate the target's position. Both TOA and TDOA require at least three TXs to determine the position [12]. The RSS technique utilizes the intensity of the signal to calculate the distance from TX to RX in determining the position [13]. These techniques achieve good performance accuracy.

TABLE I. COMPARISON BETWEEN IPS TECHNIQUES BASED ON VLC AND THE PROPOSED TECHNIQUE CSA-RSS

| Ref. | Year | Technique | Accuracy | Cost | Complexity | Additional hardware | Description |
|---|---|---|---|---|---|---|---|
| [14] | 2018 | TOA | High | High | Complex | Needed | It needs accurate synchronization between transmitter and receiver to obtain accurate estimation, which causes complexity. It needs additional hardware for synchronization, 2–3 transmitters and 2 receivers at least which causes high cost and more complexity. It has complexity in implementation. |
| [15] | 2020 | AOA | Medium | High | Complex | Needed | Its algorithm is considered complicated because of implementation and hardware. Multiple receivers and perfect estimation mechanisms are required to obtain angular information. It needs a directional receiver. |
| [16] | 2020 | RSS | Medium | Medium | Simple | Not needed | It does not require extra hardware and is cost-effective. It is less complex in implementation, but it provides medium accuracy. |
| [17] | 2021 | TDOA | High | High | Complex | Needed | It needs accurate synchronization between the transmitters to provide accurate estimation, which causes complexity. It needs additional hardware for the synchronization, 2–3 transmitters and 2 receivers at least, which causes high cost and more complexity. It has complexity in algorithm and calculation. |
| This work | 2022 | CSA-RSS | High | Low | Simple | Not needed | This proposed technique is distinguished with high accuracy based on the obtained simulation results. It is cost-effective where there is no need for additional hardware. It is simple in the calculation and implementation of positioning. |

However, there are particular limitations in using these techniques in such systems [18]. In the AOA, it needs at least two sources as transmitters to calculate the position. It requires additional hardware to measure the angle of the received signals [17]. In addition, it needs a directional receiver that is used to identify the direction from which LED light is coming [19]. These requirements add extra cost and increase the complexity. The TOA technique needs very accurate synchronization between the LEDs and the receivers [20], where it requires at least three LED references to measure the positioning [21]. It is considered complex in implementation and expensive since it requires atomic clocks [22]. The TDOA technique is also considered expensive since it needs very accurate synchronization between the LEDs. It requires at least two RXs to calculate the time difference between them [8]. Its algorithm is considered complicated since it needs to solve complicated equations compared to the TOA technique [17][20]. Despite the ease of the RSS technique implementation, it records a medium level of positioning accuracy. Its intensity follows the channel model: whenever the distance between the LED and the RX is increased, the intensity decreases, which affects the accuracy performance [9]. These techniques were focused to achieve better accuracy but neglected other important aspects such as cost and complexity. Furthermore, the case of using multiple LEDs and PDs increases cost and complexity and leads to interference occurring. Moreover, synchronization is considered complicated to implement due to the accuracy that is required. All these limitations raise the need for a technique that takes into consideration the accuracy besides cost and complexity.

Different IPS techniques based on VLC are compared with the new proposed technique CSA-RSS in terms of accuracy, cost, complexity, and additional hardware in Table I.

This paper proposes a system model for a VLC positioning system to improve the accuracy of the indoor positioning system. The contributions of this paper are summarized as follows:

- An effective indoor positioning system based on VLC technology is designed and developed to achieve high positioning accuracy of the proposed indoor system model.
- A novel technique is proposed for an indoor positioning system based on a single LED and single photodiode (PD) to exceed the limitations in the previous techniques.
- The trade-off between increasing line of sight (LOS) and different positions of the TX-RX pair is investigated to prove the effectiveness of the newly proposed technique.
- The performance parameters for different configurations in the proposed system are evaluated.

The remainder of the paper is organized as follows; Section II presents the system model. Section III demonstrates the performance of the new technique and the whole proposed system in MATLAB simulation results with the discussion. Section IV presents the conclusion.

## II. SYSTEM MODEL

The system model is shown in Fig. 1, with a typical empty room with dimensions of 5 m × 5 m × 3 m. This is the scenario in LOS of the proposed realistic VLC system model for indoor positioning in a typical room to investigate the effectiveness of the new technique. There is one LED as an access point fixed at the center of the room's ceiling in a 3D location $(X_i, Y_i, Z_i)$, and the distance between the LED and the four edges is 2.5 m. The center location of the LED is (2.5, 2.5, 3). To investigate the impact of the LED's location on the system performance, ten positions of the PD in the 3D location $(X_j, Y_j, Z_j)$, as indicated in Table II, were applied. The PD's position changes in a half-diagonal line on the floor starting from the center toward the corner as illustrated in Fig. 1. Besides the LED's functionality in illumination, it broadcasts the signals of visible light with a unique location code according to the LED's location coordinates.

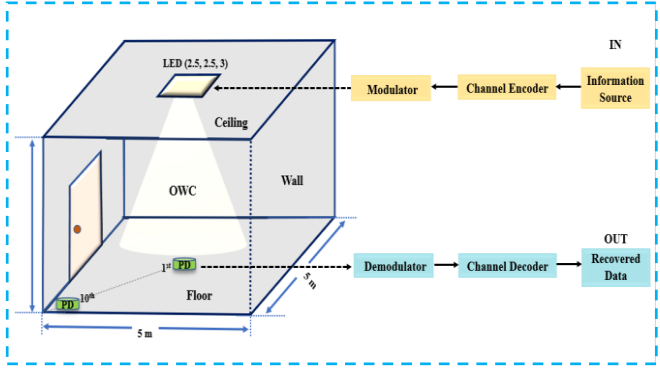

Fig. 1. Scheme of indoor VLC positioning system.

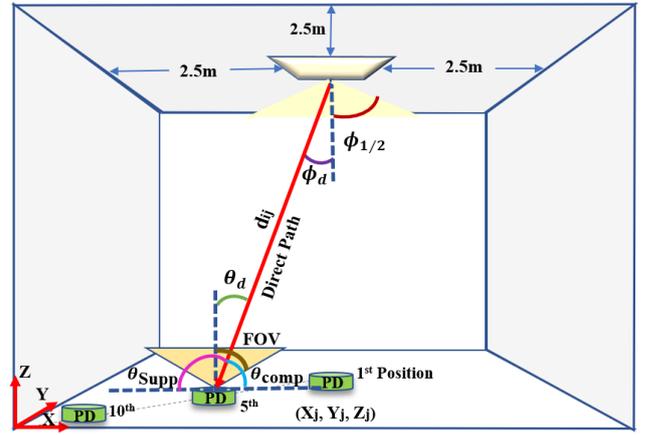

Fig. 2. Geometry of new positioning technique.

*A. Configuration of the Scenario*

- In the scenario of illumination in the indoor VLC system, a single LED as an access point for the whole room that has dimensions of 5 m × 5 m × 3 m was deployed. The LED is fixed to the room ceiling at the center with coordinates (2.5, 2.5, 3) m as described in Table II. The effect of changing the PD's location in ten positions on the floor plane needs to be investigated.

- In the first position, the PD is located under the LED directly in a straight line with the center of light's intensity, in coordinate (2.5, 2.5, 0) m, where 3 m is the distance between the LED and the PD calculated based on (7).

- The replacement of the proposed PD positions takes place symmetrically in the *X* and *Y* axes at a regular distance toward one of the corners.

- The positions change in a half-diagonal line direction on the floor to reach the last position at the corner in coordinate (0.07, 0.07, 0) m.

*B. Geometry of New Positioning Technique*

The complementary and supplementary angles based on received signal strength (CSA-RSS) technique is a novel technique proposed in this paper. The operating principle of this method is illustrated in Fig. 2, comprising all the angles required for calculation based on the LED's ray. On the PD side, the strength of the received signal is measured. The geometry model of the VLC channel between the LED and the PD in the direct path in the LOS case is shown in Fig. 2.

The distance between the LED and the four edges is 2.5 m. The position of the LED, the formed angles, and the distance in the direct path are shown in detail, where $\phi_d$ and $\theta_d$ represent the angles of irradiance and incidence respectively, with the distance $d_{ij}$ in the LOS channel. The half-power angle of the LED ($\phi_{1/2}$) and PD's field of view (*FOV*) angles have a significant role in the results. Also, the PD's locations ($X_j$, $Y_j$, $Z_j$) are shown between the center and one of the corners to investigate their impact on the performance parameters.

*1) Optical received power*

In the LOS case, the relationship between the received optical power $P_{Received}$ and the transmitted optical power $P_{trans}$ can be expressed by [23]:

$$P_{Received} = \frac{P_{trans}}{d_{ij}^2} f(\phi_d) A_{eff}(\theta_d) \qquad (1)$$

where $P_{trans}$ is the transmitted power, $f(\phi_d)$ is the radiant angle intensity at the irradiance angle $\phi_d$ with respect to the LED, $d_{ij}$ is the distance between the LED and the PD, and $\theta_d$ is the angle of incidence as illustrated in Fig. 2. $A_{eff}$ is the effective area in the PD that collects the signals and is expressed as [24]:

$$A_{eff}(\theta_d) = \begin{cases} Ah(\theta_d) g(\theta_d) \cos\theta_d, & \theta_d \leq FOV \\ 0, & \theta_d > FOV \end{cases} \qquad (2)$$

where *A* is the surface area of the PD, $h(\theta_d)$ is the optical filter gain, $g(\theta_d)$ is the concentrator gain, and *FOV* is the PD's field of view [25].

$$g(\theta_d) = \begin{cases} \frac{n^2}{\sin^2(FOV)}, & 0 \leq \theta_d \leq FOV \\ 0, & \theta_d > FOV \end{cases} \qquad (3)$$

where *n* represents the refractive index of the concentrator [26].

The radiation pattern of the LED is close to the Lambertian model. So, $f(\phi_d)$ can be expressed as [23]:

$$f(\phi_d) = [\frac{(m+1)}{2\pi}] \cos^m \phi_d \qquad (4)$$

where *m* is the Lambertian model number of an LED, and its value indicates the LED directivity. It is related to $\phi_{1/2}$, the half-power angle of the LED, as [27]:

$$m = \frac{-\ln(2)}{\ln(\cos(\phi_{1/2}))} \quad (5)$$

Then, the optical power received from a directed channel can be written as [28][29]:

$$P_{Received} = \begin{cases} P_{trans} \frac{(m+1)A}{2\pi d_{ij}^2} \cos^m(\phi_d) h(\theta_d) g(\theta_d) \cos(\theta_d), & \theta_d \leq FOV \\ 0, & \theta_d > FOV \end{cases} \quad (6)$$

The distance between the LED and the PD in the LOS channel can be calculated as [29][30]:

$$d_{ij} = \sqrt{(X_i - X_j)^2 + (Y_i - Y_j)^2 + (Z_i - Z_j)^2} \quad (7)$$

The horizontal distance ($D_{hor}$) between the LED and the PD can be obtained as follows [22][31]:

$$D_{hor} = \sqrt{d_{ij}^2 - V^2} \quad (8)$$

*2) Positioning error based on new technique CSA-RSS*

The positioning error is considered one of the significant performance parameters in the VLC system. In order to find the estimated position of the PD, angulation calculation by the novel CSA-RSS method is used. This method depends on calculating the complementary angle to the angle of incidence obtained from simulation results. Then, the supplementary angle is calculated to the complementary angle based on the angle and incidence ray where their sum equals to 180°. Hence, to illustrate the relationship between the complementary and supplementary angles with incidence angle $\theta_d$, it can be described as follows:

$$\left.\begin{array}{l} \angle\theta_d + \angle\theta_{comp} = 90° \\ \angle\theta_{comp} = 90° - \angle\theta_d \\ \angle\theta_{supp} = 90° + \angle\theta_d \\ \angle\theta_{comp} + \angle\theta_{supp} = 180° \end{array}\right\} \quad (9)$$

Subsequently, by using the trigonometry functions, ($X$, $Y$) points for the complementary and supplementary angles have been calculated as follows:

$$\left.\begin{array}{l} X_{comp} = D_{hor} \cos(\theta_{comp}) \\ Y_{comp} = D_{hor} \cos(\theta_{comp}) \end{array}\right\} \quad (10)$$

$$\left.\begin{array}{l} X_{supp} = D_{hor} (\sin\theta_{supp}) \\ Y_{supp} = D_{hor} (\sin\theta_{supp}) \end{array}\right\} \quad (11)$$

The average of the obtained coordinates from the CSA-RSS technique refers to the PD's estimated position in different locations which are represented as $X'$ and $Y'$ while $Z' = 0$ because the PD is moving on the floor in all positions, and the equations are as follows:

$$\left.\begin{array}{l} X' = \dfrac{X_{comp} + X_{supp}}{2} \\ Y' = \dfrac{Y_{comp} + Y_{supp}}{2} \end{array}\right\} \quad (12)$$

Then, to obtain the positioning error, subtracting the PD's actual position from the estimated one that is obtained from the CSA-RSS method is given by [32]:

$$\text{Positioning error} = \sqrt{(X_j - X')^2 + (Y_j - Y')^2 + (Z_j - Z')^2} \quad (13)$$

The system configuration and simulation parameters values that are considered in this paper are given in Table II.

III. SIMULATION RESULTS AND DISCUSSION

This section summarizes the simulation findings that investigate an efficient indoor positioning system based on VLC to improve the accuracy of position error. The simulation's parameters are listed in Table II. The model is proposed to analyze the performance of the LOS scenario for different configurations of PD position. The aim is to prove the effectiveness of the newly proposed technique with increased requests to acquire high-speed transmission data with reduced cost. The distribution scenario is for a single LED and ten different positions of the PD in a 5 m × 5 m × 3 m room environment. The LED is fixed at the center of the room's ceiling. The PD has ten different positions changing in a half-diagonal line on the floor starting from the room's center under the LED directly in a straight line toward one of the corners. The replacement of the proposed PD positions takes place symmetrically in the $X$ and $Y$ axes to enable the evaluation of the performance regularly.

TABLE II. SYSTEM SIMULATION PARAMETERS

| Parameter | Value |
|---|---|
| Room dimensions | 5m × 5 m × 3 m (width × length × height) |
| No. of LEDs | 1 |
| LED location ($X_i$, $Y_i$, $Z_i$) (m) | (2.5, 2.5, 3) |
| No. of PDs (receiver) | 1 |
| PD's ten locations ($X_j$, $Y_j$, $Z_j$) (m) | (2.50, 2.50, 0), (2.23, 2.23, 0) (1.96, 1.96, 0), (1.69, 1.69, 0) (1.42, 1.42, 0), (1.15, 1.15, 0) (0.88, 0.88, 0), (0.61, 0.61, 0) (0.34, 0.34, 0), (0.07, 0.07, 0) |
| Surface area of the PD | 2.25 mm$^2$ |
| Transmitted power from LED | (8, 10, 12, 15) watts |
| Optical filter gain $h(\theta_d)$ | 1.0 |
| Reflective index $n$ | 1.5 |
| Lambertian model number $m$ | 1.3 |
| Half-power angle of the LED | 60° |
| Field of view FOV | 90° |
| Incidence angle $\theta_d$ | 60°, 70°, 80°, 90° |

*A. Performance of Received Optical Power*

The received power is a crucial parameter in evaluating any proposed system. Receiving a robust signal at the PD side is considered a good indicator of system performance. To examine the performance of the received power with design parameters according to (6), varying values of incidence angles and transmitted power were chosen to verify its performance and robustness.

Fig. 3 shows the performance of received power performance versus distance with different incidence angles of 60°, 70°, 80°, and 90°. The received power value impacted the distance and the varying incidence angles values because of the difference in the distance between the

LED source and the ten positions of the PD. Therefore, incidence angles are considered one of the important parameters to VLC system design based on (6). Hence, the performance of received power decreases while the distance between the LED and the PD increases. For example, the maximum received power values are achieved in the first position of the PD in the case of an LED-PD distance of 3 m. It is a fact due to the short distance between the LED source and the PD destination. On the other side, the minimum received power values are achieved at the tenth position of the PD, where the LED-PD distance is 4.56 m, due to the far LED source effect in received power to the PD. As a result, the incidence angle has a key role in the achievable performance. For example, the maximum value of received power is achieved at 90° due to stronger LOS with the center of LED source intensity. On the other side, the minimum received power value was within 0.77–0.34 W at the incidence angle of 60°. In the other considerations, for the case of incidence angles of 70° and 80°, the performance of the received power value shows a decrement of 1.53–0.67 W and 3.01–1.32 W, respectively. In conclusion, the maximum received power obtained values at 60° and 90° are obtained at 3 m, and the difference between them is 3.73 W. From the obtained results, it is noticeable that the received power decreases versus the LED-PD distances simultaneously increasing due to being far from the LOS channel. The highest received power is achieved in this relationship at 3 m and 90° with a value of 4.5 W.

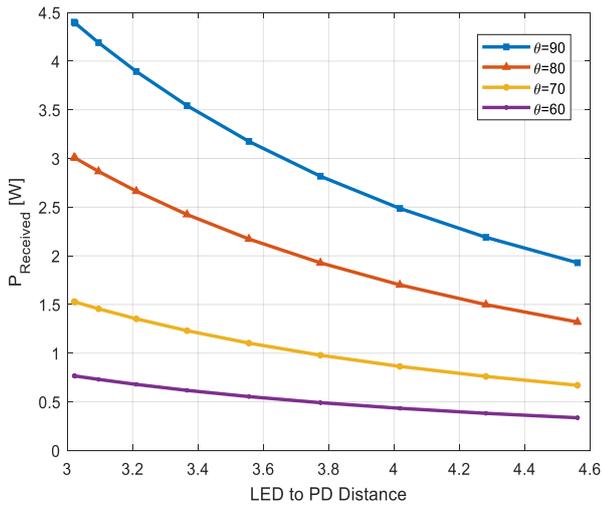

Fig. 3. Performance of received power versus distance between the LED-PD with different incidence angles.

Fig. 4 shows the received power performance versus the distance between the LED and different positions of the PD with varying values of transmitted power of 8, 10, 12, and 15 W. The performance of received power decreased with increasing distances for varying transmitted powers. In this context, increases in the distance during the PD movement toward the corner act to a decrease in the received power simultaneously. It is noticeable that the maximum received power values are achieved at a distance of 3 m between LED-PD, which represents the first position of the PD. The minimum received power values were achieved at a distance of 4.56 m between LED-PD, which represents the tenth position of the PD near the corner. The higher value of received power is achieved at the transmitted power of 15 W due to stronger transmission power from the LED source leading to an increase in the LOS in the destination to the PD.

In this context, the received power decreased from 2.34 to 1.02 W in the case of transmitted power 8 W. In the other consideration for the cases of transmission power of 10 and 12 W, the received power value increased to reach 2.92 W and 3.51 W, respectively. With the PD's movement toward the corner and increased distance, its value decreased to reach 1.29 W and 1.54 W respectively. The maximum received power value is achieved in the case of applying the highest transmitted power value for the LED source, which is 15 W. Meanwhile, the received power value increases to 4.5 W versus the LED-PD distance of 3m in the initial stage. After that, in the case of LED-PD distance increasing the received power significant decreased to reach 1.92 W due to weak received power with distance.

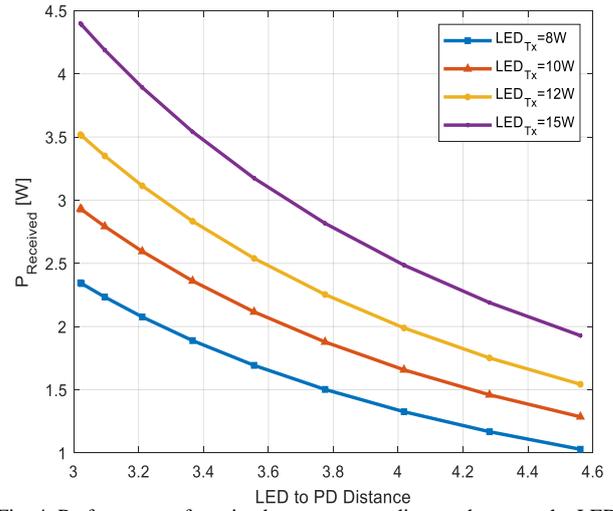

Fig. 4. Performance of received power versus distance between the LED-PD with different values of transmitted power.

*B. Positioning Error Based on New Technique CSA-RSS*

Indoor positioning becomes a crucial part of smart living and advanced technology in the era of the internet of things (IoT). Hence, to prove the effectiveness of the new proposed technique CSA-RSS as a solution for determining positioning accurately, a performance examination was made as described in Fig. 5 (a) and (b).

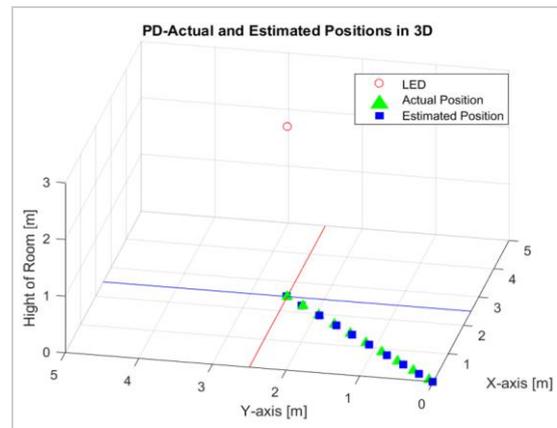

(a)

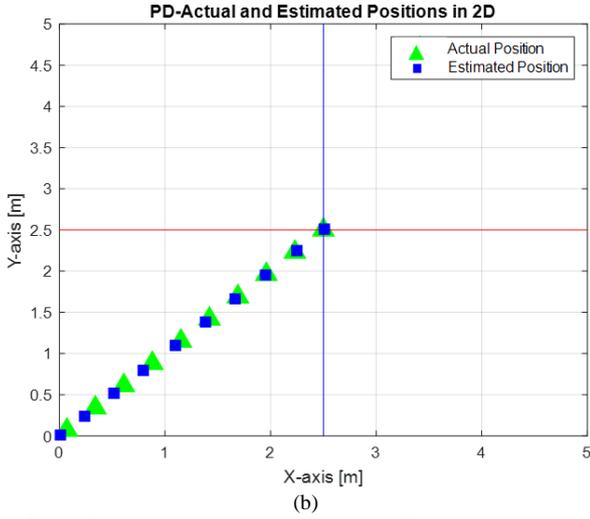

Fig. 5. (a) 3D actual and estimated positions of PD; (b) describes the plane of the PD in 2D for the room's floor (5m×5m), where the length and the width of the floor have been divided to show the center location.

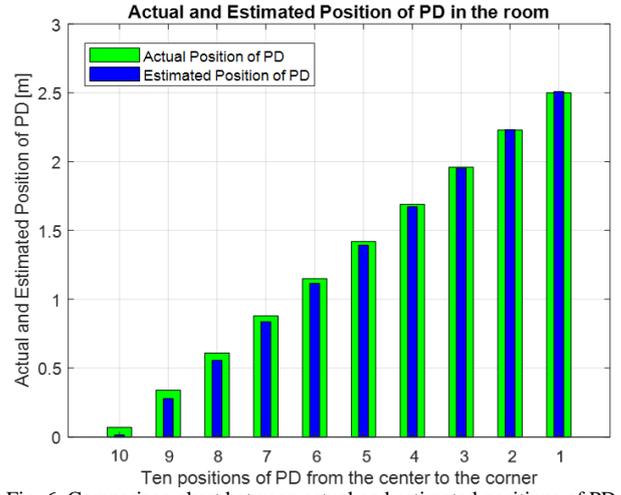

Fig. 6. Comparison chart between actual and estimated positions of PD.

The actual position is represented as the green triangles while the estimated position is represented as the blue squares. Hence the CSA-RSS technique calculated the estimated positions of the PD according to (9)–(12). It can be observed that there is great convergence between the actual and the estimated position in the room's center at 3 m and a decrease gradually whenever PD moved toward the corner. A better result was obtained in the first position when the PD is in direct LOS with the LED at a distance of 3 m, where the angles of the CSA-RSS technique are equal to 90°. Whenever the distance is increased between the LED and the PD, the performance level will be less. This is attributed to the LED source being closer to the PD destination. It is noticeable that the estimated positions have symmetrical values on the *X* and *Y* axes for each location and it is similar to the symmetry of the actual coordinates. This is because the movement takes place on the half-diagonal line toward the chosen corner.

To examine the performance effectiveness of the CSA-RSS technique, a comparison between the actual and the estimated position is represented as bars in Fig. 6 according to their coordinates. The green bars represent the actual fixed known positions of the PD while the blue bars represent the estimated position obtained based on CSA-RSS in the chosen floor quarter, where the PD changes its location. The results of the bar comparison show a slight variance between the actual and the estimated position in the room's center at a direct line with the LED. The actual coordinates for the first position show great convergence.

By examining the effect of moving the PD along the half-diagonal line toward the corner on the estimated positions, it shows a decrease in the accuracy with increasing the positioning error values. The estimated and actual coordinates for the ten positions are illustrated in Table III. The last estimated position at the corner shows a bigger difference than the others because of the increase in the distance far from the LED and the LOS and is affected more by channel factors.

In Fig. 7, the positioning error of the proposed system for ten positions is represented in blue bars. The *X*-axis represents the PD position arrangement from 1 to 10 while the *Y*-axis represents the positioning error values in meters.

The obtained results are calculated based on (13). The positioning error for the first position is 0.0013 m, which indicates a significant convergence to the actual one. The values of the positioning error increase versus the distances simultaneously increasing far from the LED source. Therefore, the positioning error at the tenth position witnesses a significant increase to reach 0.0797 m. In this context, the difference between the first and the tenth position is 0.0784, which means it increases by 7.84%.

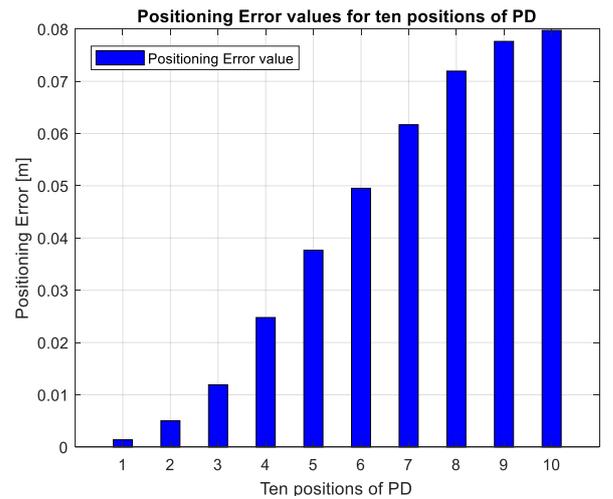

Fig. 7. Positioning error of ten positions of PD based on CSA-RSS.

TABLE III. POSITIONING ERROR BASED ON ACTUAL AND ESTIMATED POSITIONS OF PD

| Position | PD's actual coordinates | PD's estimated coordinates | Positioning error (m) | Average positioning error (m) |
|---|---|---|---|---|
| 1st (at center) | 2.5, 2.5, 0 | 2.5009, 2.5009, 0 | 0.0013 | 0.042 m |
| 2nd | 2.23, 2.23, 0 | 2.2336, 2.2336, 0 | 0.0050 | |
| 3rd | 1.96, 1.96, 0 | 1.9515, 1.9515, 0 | 0.0118 | |
| 4th | 1.69, 1.69, 0 | 1.6724, 1.6724, 0 | 0.0247 | |
| 5th | 1.42, 1.42, 0 | 1.3933, 1.3933, 0 | 0.0376 | |
| 6th | 1.15, 1.15, 0 | 1.1149, 1.1149, 0 | 0.0495 | |
| 7th | 0.88, 0.88, 0 | 0.8363, 0.8363, 0 | 0.0616 | |
| 8th | 0.61, 0.61, 0 | 0.5591, 0.5519, 0 | 0.0719 | |
| 9th | 0.34, 0.34, 0 | 0.2851, 0.2851, 0 | 0.0776 | |
| 10th (at corner) | 0.07, 0.07, 0 | 0.0136, 0.0136, 0 | 0.0797 | |

Table III illustrates the actual and estimated ten coordinates of the PD with their positioning error, which represent the difference between them according to (13). The average of the positioning error is a good indicator for the proposed system, where the achieved result is 0.042 m, proving the success of this technique in indoor positioning with fewer requirements and less complexity compared with other techniques.

## IV. CONCLUSION

VLC systems are gaining so much interest from industry, government, and academia because it is considered as a root solution that suits the rapid expansion occurring in the communication and technologies world. Furthermore, many techniques applied with VLC for indoor positioning achieved a good level in terms of accuracy. However, some of them need additional hardware devices that cause extra cost, and some need synchronization which increase the complexity.

This paper considers achieving high accuracy, reduced cost, less complexity, and efficiency simultaneously by designing and developing an efficient indoor positioning system based on VLC technology. Investigation and analysis have been done for different configurations in terms of positioning error and received power. The simulation results of the proposed system demonstrate its effectiveness with good and reasonable outcomes obtained from MATLAB simulation. It achieved high accuracy in terms of low positioning error in locating the PD position, where the average error is 3.2 cm in 80% of the examined positions and the maximum average positioning error is 4.2 cm.